\documentstyle[epsfig]{mn}

\title[A completeness test for  
magnitude-redshift
samples]{A simple tool for assessing the completeness in apparent
magnitude  
of magnitude-redshift
samples}
\author[S. Rauzy]
       {St\'ephane Rauzy\\
        Department of Physics and Astronomy, University of Glasgow,
Glasgow, G12 8QQ, UK.}
 \date{Accepted .
      Received ;
      in original form }

\begin{document}

\maketitle

 \begin{abstract}
   
A new tool is proposed for finding out the completeness limit
in apparent magnitude 
of a magnitude-redshift sample. The technique, closely related to 
the statistical test proposed by Efron \& Petrosian (1992),
 presents a real improvement
compared to standard
completeness tests. Namely, no a priori assumptions are required 
concerning the 
redshift space distribution of the sources. It means in particular
that neither the clustering nor the evolution of the mean number density
of the galaxies
do affect the result of the search.
 \end{abstract}


\section{Introduction}
\label{Introduction}

Extracting the intrinsic characteristics of the galaxies population
from magnitude-redshift samples (e.g. the luminosity function,
the power spectrum of the spatial density fluctuations) 
remains one of the major concerns
of observational cosmology. The task is somehow complicated
due to the presence of selection effects in observation, e.g.
a detection threshold in apparent fluxes.
Because a part of the population is indeed not observed, standard
statistical methods lead in general to biased estimate of
the genuine parameters characterizing the population.
While correcting for such biases is hardly feasible
when intricate selection effects are at work,
the problem has been fortunately handled in some special
cases.

Samples complete in apparent magnitude obviously deserve mention.
Indeed, magnitude-redshift data truncated to a lower flux limit
has been extensively studied in the literature.
Powerful methods have been developed in such a case for fitting 
or reconstructing
the luminosity function of the galaxies population
(for example the $C^-$ method of Lynden-Bell 1976, the maximum likelihood
fitting technique of Sandage et al. 1979). If the sample is
furthermore complete in redshift, sophisticated methods have been proposed
for estimating the power spectrum of the galaxies distribution
(e.g.  Heavens \& Taylor 1997) and for inferring the cosmic velocity
field from magnitude-redshift data (e.g. Rauzy \& Hendry 2000,
Branchini et al.  1999). 
 
Flux limited magnitude-redshift samples can be built 
by primarily  selecting from a parent magnitude sample all the objects 
brighter than the adopted flux limit. In a second step is 
collected the redshift of each of these sources (in this case the sample
is complete in redshift), or a random subselection of these galaxies. 
By following such a strategy of observation, the completeness in apparent 
magnitude is essentially warranted as long as the parent sample is
itself complete up to the flux limit. However, 
the completeness of the parent sample is in general difficult to
assess. Undesired selection effects in observation are often at work
and to some extent, the selection process depends on how are defined
and measured the
magnitudes of the galaxies (e.g. isophotal, visual, total magnitudes), how
surface brightness threshold  affects the sample, etc... (see for example
Sandage \& Perelmuter 1990, Petrosian 1976, Driver 1999). Moreover,
the parent sample may have been selected using criteria involving
other observables not straightforwardly related to magnitudes 
(e.g. diameters, fluxes in a different passband).
It turns out that the completeness assumption, a crucial prerequisite 
for applying any fitting and recontruction methods mentioned above,
must be checked thoroughly.

A classical test for completeness in apparent magnitude is to
analyse the variation of the galaxies number counts in function of the limiting
apparent magnitude (Hubble 1926). This test, which presupposes
that the galaxies population does not evolve with time and
is homogeneously distributed
in space, is however not very efficient. It is
difficult to decide in pratice whether the deviation of the counts law is due
to the presence of clustering and evolution of the galaxies
luminosity function, or is indeed created by incompleteness in apparent
magnitude. Including the redshift information, the $V/V_{\rm max}$ test
of Schimdt 
(1968) has been also used for assessing 
the completeness of magnitude-redshift samples (see for example
Hudson \& Lynden-Bell 1991), but suffers unfortunately from the
same major drawback than the Hubble completeness test. 
  
Efron \& Petrosian (1992) have thoroughly analysed the statistical
properties of magnitude-redshift samples complete in apparent magnitude.
They proposed a robust test for independence, free of assumption 
concerning the spatial distribution of sources, which allows to
estimate the cosmological parameters characterizing the geometry of the
Universe from a quasar sample (see also Efron \& Petrosian 1999). 
It turns out that this statistical test, given a world model, can easily
be recycled for testing the completeness assumption. The purpose of
the present paper is exactly to take advantage of such a possibility. 

The statistical background of the method as well as the test for
completeness are presented section 2. An example of application
is given in section 3, where the test is used for investigating the 
completeness in apparent
magnitude of the South Sky Redshift Survey  of da Costa et al. (1998).
The properties of the new completeness test and its range
of application are finally summarised in section 4.  

\section{The completeness test}

\subsection{Assumptions and statistical model}

The luminosity function of the galaxies population is herein defined
following Bingelli et al. (1988) as the probability distribution
function $f_t(M)$ of the absolute magnitude $M$ of the galaxies depending
in general on the epoch $t$. At any epoch, 
the luminosity function is by definition normalised (i.e. 
$\int f_t(M)\,dM =1$).
It is assumed hereafter that the
luminosity function of the population does not depend on the
3D redshift space position ${\bf z}=(z,l,b)$ of the galaxies. Without accounting
for selection effects in observation, the probability density
describing the population splits under this assumption as
\begin{equation}\label{dP_1}
dP_{{\bf z}M} \propto dP_{\bf z} \times dP_M= \rho(z,l,b)\,
dl db dz \times
f(M) dM
\end{equation}
where $\rho(z,l,b)$ is the 3D redshift space distribution function of the
sources
along the past light-cone. 

The present model is thus well suited to describe
the observed spatial fluctuations of the galaxies density and
to account for a pure density
evolution scenario (i.e. the variations of the mean galaxies density with
redshift or equivalently with time). 
On the other hand, Eq. (\ref{dP_1}) fails to describe
environmental effects (i.e. the luminosity function of the sampled
objects depends on the local environment) and an
evolution of the specific characteristics (e.g. mean absolute magnitude, shape)
of the luminosity function.

Selection effects in observation enter the statistical model
as a filter response function (see for example Bigot\&Triay
1990). This selection function $\psi$ can be expressed in general in terms of
the observable quantities, namely 
herein the line-of-sight direction $(l,b)$, the redshift $z$
and the raw apparent
magnitude $m$, i.e.
$\psi \equiv \psi(m,z,l,b)$. Accounting for selection effects in
observation, the probability density describing the sample may
be written as
\begin{equation}\label{dP_2}
dP = \frac{1}{A}\psi(m,z,l,b)\,\rho(z,l,b)\,f(M)
dl db dz 
dM
\end{equation}
with $A$ the normalisation
factor warranting $\int dP =1$.

The null hypothesis tested hereafter is that the sample is
complete in raw apparent magnitude
up to
a given magnitude limit $m_{\rm lim}$, or in other words where the selection
function in apparent magnitude is well described by a sharp cut-off, i.e.
\begin{equation}\label{Selectioneffects}
{\rm H0:}\,\,\,\,\psi(m,z,l,b) \equiv 
\theta(m_{\rm lim}-m)
\times
\phi(z,l,b) 
\end{equation}
with $\theta(x)$ the Heaviside or `step' function. The function 
$\phi(z,l,b)$ describes some eventual selection effects in angular
position and observed redshift. For example, it could account for
a mask in angular position as well as pure selection or subsampling 
in redshift
(e.g. a lower and upper limits).

The absolute magnitude $M$ is  obtained following 
\begin{equation}\label{M}
M = m_{\rm cor} - \mu(z) 
\end{equation}
where the distance modulus $\mu(z)$ can be evaluated from the
redshift given a cosmological world model $\{H_o,\Omega_o,\Lambda_o \}$
(see for example Weinberg 1972). Note that it has been implicitly
assumed herein that the contribution of peculiar velocities to the 
observed redshifts is negligible.
The corrected apparent magnitude 
$m_{\rm cor}$ is expressed as 
\begin{equation}\label{mcor}
m_{\rm cor}= m - k_{\rm cor}(z) - A_g(l,b) 
\end{equation}
with  
$k_{\rm cor}(z)$ standing for a k-correction term  and $A_g(l,b)$
accounting for a Galactic extinction correction. 

At this stage, it is convenient to introduce the 
quantity $Z$ defined as 
\begin{equation}\label{Z}
Z = m - M = 
\mu(z) + k_{\rm cor}(z) + A_g(l,b) 
\end{equation}
which can be computed from 
the observables $z$ and $(l,b)$ providing 
a world model.  
Under the null hypothesis of Eq. (\ref{Selectioneffects}),
i.e. the sample is complete in apparent magnitude, 
the probability density of Eq. ({\ref{dP_2}) may
be rewritten using these notations as
\begin{equation}\label{dP_3}
dP = \frac{1}{A} h(Z,l,b) \,dl db dZ \, f(M) dM
\, \theta(m_{\rm lim}-m)
\end{equation}
where the distribution function $h(Z,l,b)$ may be expressed,
if required, in function of the 3D redshift space distribution
$\rho(z,l,b)$, 
the selection function 
$\phi(z,l,b)$ introduced Eq. (\ref{Selectioneffects}) and
using the definition of $Z$ given Eq. (\ref{Z}). The
cut-off in apparent magnitude introduces a correlation between
the variables $M$ and $Z$ (intrinsically faint and distant galaxies
are discarded, see figure 1) wich would have been statistically
independent otherwise.

\subsection{The random variable $\zeta$}

Thanks to the introduction of the quantity $Z$, 
the maximum absolute magnitude 
$M_{\rm lim}(Z)$
for which
a galaxy at a given $Z$ would be visible in the sample
is uniquely defined, i.e.
\begin{equation}\label{Mlim}
M_{\rm lim}(Z)=m_{\rm lim} - Z
\end{equation}
The milestone of the method consists in defining  the random
variable $\zeta$ as follows
\begin{equation}\label{zeta}
\zeta = \frac{F(M)}{F(M_{\rm lim}(Z))}
\end{equation}
where $F(M)$ stands for the Cumulative
Luminosity Function, i.e.
\begin{equation}\label{CLF}
 F(M)=\int_{-\infty}^M f(x)dx 
\end{equation}
The volume element of Eq. (\ref{dP_3}) may thus be rewritten as
\begin{equation}\label{dzetadmu}
dl db dZ \,d\zeta = \frac{f(M)}{F(M_{\rm lim}(Z))} \,dl db dZ \,dM
\end{equation}
and by definition the random variable $\zeta$ for a
sampled galaxy belongs to the interval $[0,1]$. The probability
density of Eq. (\ref{dP_3}) reduces therefore to
\begin{equation}\label{dP_4}
dP =  \frac{1}{A} h(Z,l,b)\,F(M_{\rm lim}(Z))
\,dl db dZ  \times 
\theta(\zeta)
 \theta(1-\zeta)
\, d\zeta
\end{equation}
with $A= \int h(Z,l,b)\,F(M_{\rm lim}(Z)) \,dl db dZ$. 
It follows from Eq. (\ref{dP_4}) that:
\begin{itemize}
\item P1: $\zeta$ is uniformly distributed between $0$ and $1$.
\item P2: $\zeta$ and $(Z,l,b)$ are statistically independent.
\end{itemize}
Property P1 will be used hereafter to construct the test for completeness.

\begin{figure}
\vbox
{\epsfig{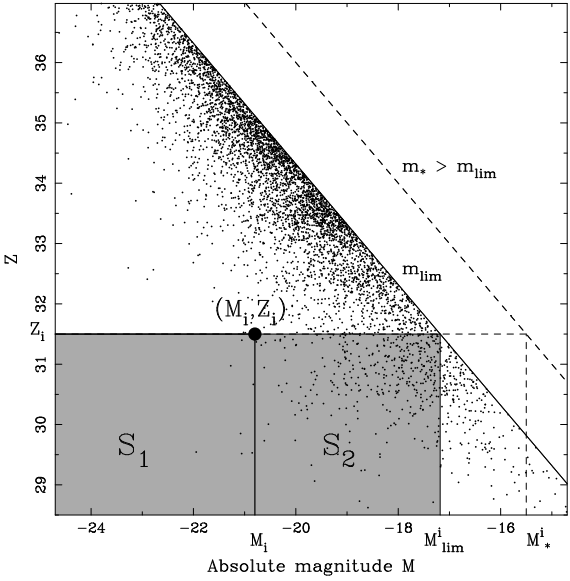}}
\caption{
The variable $Z$ defined Eq. (6) versus absolute magnitude $M$.
The procedure for evaluating $r_i$ and $n_i$ entering the 
calculation of the random variable ${\hat \zeta_i}$ of 
Eq. (15) is illustrated.
}
\label{figure1}
\end{figure}

\subsubsection{Estimate of the random variable $\zeta$}

Under the null hypothesis H0, the random variable $\zeta$ 
can be estimated without any prior
knowledge of the cumulative luminosity function $F(M)$. Let us consider
the distribution of the sampled galaxies in the $M$-$Z$ diagram
(see figure 1).
To each point with coordinates $(M_i,Z_i)$ is associated the region 
$S_i=S_1 \cup S_2$ defined as 
\begin{itemize}
\item $S_1 =  \{(M,Z) {\rm ~such~that~}
 M \le M_i {\rm ~and~} Z \le Z_i \}$
\item $S_2 =  \{(M,Z) {\rm ~such~that~} M_i <  M \le M^i_{\rm lim}
 {\rm ~and~} Z \le Z_i \}$
\end{itemize}
The random variables $M$ and $Z$ are independent in each subsample $S_i$ 
since by construction the cut-off in apparent magnitude is superseded
by the constraints 
$ M \le M^i_{\rm lim}(Z_i)$ and $Z \le Z_i$
(see figure 1). 
It implies from Eq. (\ref{dP_3}) that the number of points $r_i$ 
belonging to $S_1$ 
is given by 
\begin{equation}\label{r_i}
\frac{r_i}{N_{\rm gal}} =F(M_i) \,\times \,
\frac{1}{A}\,\int_{-\infty}^{Z_i} h(Z,l,b)\,dZdldb
\end{equation}
with $N_{\rm gal}$ the number of galaxies in the sample, and that
the number of
points $n_i$ in 
$S_i=S_1 \cup S_2$ is 
\begin{equation}\label{n_i}
\frac{n_i}{N_{\rm gal}} =F(M_{\rm lim}(Z_i)) \,\times \,
\frac{1}{A}\,\int_{-\infty}^{Z_i} h(Z,l,b)\,dZdldb
\end{equation}
The numbers $r_i$ and $n_i$ are obtained by merely counting the
galaxies respectively belonging to $S_1$ and $S_1 \cup S_2$. 
An unbiased estimate of the
random variable $\zeta$ introduced Eq. (\ref{zeta}) is indeed provided
by the quantity 
\begin{equation}\label{zetaestimate}
{\hat \zeta_i} =  \frac{r_i}{n_i+1}
\end{equation}
Using rank-based statistics, Efron \& Petrosian (1992) prove
moreover that the random variables   
${\hat \zeta_i}$ are independent of each other under H0.
The expectation $E_i$ and variance $V_i$ of the  ${\hat \zeta_i}$ 
are respectively
\begin{equation}\label{Expectationandvariance}
E_i = \frac{1}{2} 
\,\,\,\,\,\,\,
\,\,\,\,\,\,\,;
\,\,\,\,\,\,\,
\,\,\,\,\,\,\,
V_i = \frac{1}{12} \,
\frac{n_i-1}{n_i+1}
\end{equation}
Note that the value of the variance $V_i$ tends towards the variance of
a continuous uniform distribution between 0 and 1 when $n_i$ 
becomes large
enough. 

It turns out that the quantity $T_C$ defined as
\begin{equation}\label{T_C}
T_C = 
\left .
{
\displaystyle
 \sum_{i=1}^{N_{\rm gal}} 
\left ( {\hat \zeta_i} - \frac{1}{2} \right )
} 
\right /
{
\displaystyle
\left (\sum_{i=1}^{N_{\rm gal}} V_i \right )^{\frac{1}{2}}
} 
\end{equation}
has an expectation zero and variance unity under H0. The statistic $T_C$
proposed herein is almost similar to the Efron \& Petrosian test statistic
for 
independence
based on the normalised ranks.

The $T_C$ statistic can be estimated without assuming any prior model
for the luminosity function. It is worthwhile to mention that no
assumptions have been made as well concerning the distribution function
$h(Z,l,b)$ introduced Eq. (\ref{dP_3}). It means that the property
of the $T_C$ statistic derived above holds for any 
3D redshift space distribution
$\rho(z,l,b)$ (allowing the presence of clustering and the evolution of
the mean galaxies density with time), and for any 
selection function 
$\phi(z,l,b)$ (e.g. subsampling in redshift bins would not bias the
$T_C$ statistic).

\begin{figure}
\vbox
{\epsfig{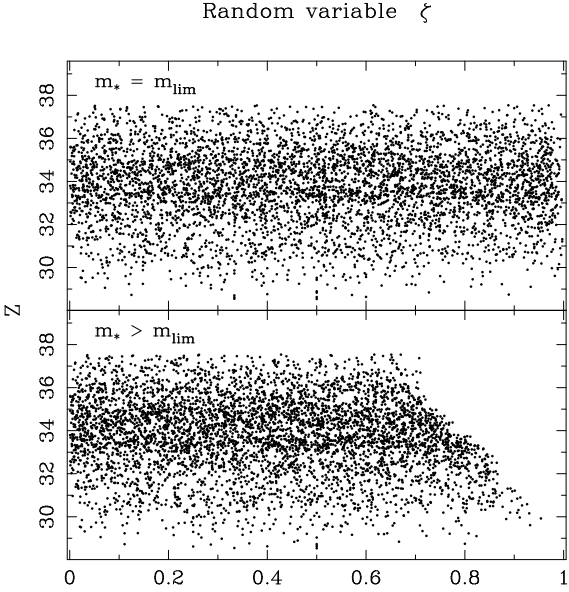}}
\caption{
Diagram $\zeta$-$Z$ for two values of the limiting apparent magnitude
$m_*$. For $m_*$ greater than the completeness limit $m_{\rm lim}$, 
the number of galaxies fainter than $M_{\rm lim}(Z)$ is underestimated
due to incompleteness (see figure 1), inferring a systematic lack of 
points with a value of ${\hat \zeta_i}$ close to unity.
This effect is particularly visible at high $Z$ (i.e. distant galaxies).
}
\label{figure2}
\end{figure}

\subsection{The test for completeness $T_C$}

The principle of the test is to evaluate the quantity 
$T_C$ defined Eq. (\ref{T_C}) for subsamples truncated 
to increasing apparent magnitude limit $m_*$, i.e. 
$m_*$ is replacing $m_{\rm lim}$ in Eq. (\ref{Mlim}). 

As long as $m_*$ remains 
below the completeness limit $m_{\rm lim}$ the subsample is obviously
complete up to $m_*$ and the $T_C$ statistic
is thus expected to be distributed around zero with sampling fluctuations
of dispersion 
of unity. 
On the other hand as $m_*$ becomes greater than $m_{\rm lim}$,
the incompleteness introduces a deficit of galaxies with $M$
fainter than $M_{\rm lim}(Z)$
(see figure 1). It results in a lack of galaxies
with a value of ${\hat \zeta_i}$ close to unity. 
Figure 2 dramatically illustrates such a trend
for a sample characterized by a strict cut-off in 
$m_{\rm lim}$, but the trend would remain similar for a smooth
imcompleteness function as well.
It turns out that the $T_C$ statistic is expected to
be systematically negative for
limiting apparent magnitude $m_*$ greater than the completeness
limit $m_{\rm lim}$. 

Therefore, the curve $T_C(m_*)$ is
characterized by a plateau of zero mean
for $m_*$ below 
$m_{\rm lim}$, followed
by a systematic decline beyond this apparent magnitude, 
i.e.
\begin{equation}\label{T_C_Test}
T_C \simeq 0~~{\rm for}~~m_* \le m_{\rm lim}~~;~~
T_C < 0~~{\rm for}~~m_* > m_{\rm lim}
\end{equation}
Under H0, the sampling fluctuations make the $T_C$ statistic to follow 
a gaussian distribution of
variance one. The decision rule for fixing the completeness limit is a matter
of choice, having in mind that the confidence levels of rejection 
associated to the events 
$T_C < -1$,
$T_C < -2$,  
$T_C < -3$
are respectively 
$84.13\%$,
$97.72\%$, 
$99.38\%$ (i.e. these numbers correspond to the probabilities drawn
from a Normal distribution for a 
one-sided rejection test). 

\section{Example of application}

The test for completeness $T_C$ is herein applied to the 
South Sky Redshift Survey (SSRS2) sample
of da Costa et al. (1998). The sample, containing $5369$ galaxies
with measured B-band magnitude and redshift, has been drawn primarily
from the list of nonstellar objects identified in the Hubble Space
Telescope Guide Star Catalog (Lasker et al. 1990). 
The redshift survey is more than
$99\%$ complete up to  
the magnitude limit $m_{\rm SSRS2}$ of 15.5 mag (da Costa et
al. 1998). 
The test for completeness $T_C$ is thus applied to assess the completeness
in apparent magnitude of the primary list of galaxies used
to build on the SSSR2 sample.

The type-dependent k-correction are calculated following Pence (1976),
i.e.
$k_{\rm cor}(z) = K_B(T) \times cz/(10\,000~{\rm km~s}^{-1})$ with
$K_B(T)=0.15$ for $T \le 0$,  
$K_B(T)=0.15 -0.025\,T$ for $3 \ge T \ge 0$ and  
$K_B(T)=0.075 -0.010\,(T-3)$ for $3 \le T $.
Galactic extinctions are obtained as $A_g(l,b) = 4.325\,E(B-V)$
by use of the dust maps of Schlegel et al. (1998) for the
redenning correction.

The redshifts have been transformed in the CMB rest frame and
the distance modulus is computed adopting an Hubble constant of
$H_0=100$ km s$^{-1}$ Mpc$^{-1}$ in a flat universe with no
cosmological constant (i.e. $\Omega_0=1$ and $\Lambda_0=0$).  

Galaxies not belonging to the redshifts range $[2\,500,15\,000]$
km s$^{-1}$ are discarded. The lower limit in redshift
has been introduced in order to minimize the impact of peculiar
velocities on the $T_C$ statistic. In particular, the kinematical 
influence of the Virgo cluster will be considerably reduced by removing
nearby galaxies. 
The upper bound in redshift sets some limits on the interval of time
spanned by the data and thus reduces the influence of an eventual evolution
of the luminosity function on the $T_C$ statistic 
($cz =15\,000$ km s$^{-1}$ corresponds to a look-back time of
$0.5\,10^9$ years for $H_0=100$ km s$^{-1}$ Mpc$^{-1}$, respectively 
$1$ Gyr if $H_0=50$ km s$^{-1}$ Mpc$^{-1}$).
Note that such a subsampling in redshift is not expected to
affect the result of the search. 

Finally, 
a random component uniformly distributed between $[-0.005,0.005]$  
has been added to the catalogued magnitudes 
(the apparent magnitudes of the SSRS2 sample 
are rounded to $0.01$ mag).
Rounding problems have negligible effects on the
present analysis but can infer spurious variations of the
$T_C$ statistic. Because magnitudes are 
set to discrete values, it creates some artificial gaps in the magnitude
distribution function. The effect is  observable
on datasets rounded high and containing a large number of galaxies,
e.g. 
the Zwicky catalogue (Zwicky et al. 1968).

\begin{figure}
\vbox
{\epsfig{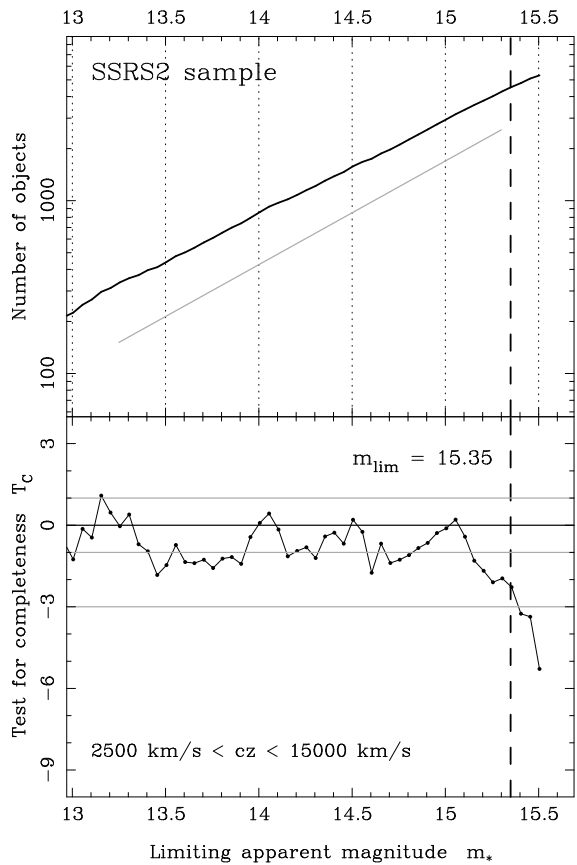}}
\caption{The test for completeness $T_C$ applied to the
the $4324$ galaxies (all types) of the SSRS2 sample with redshifts
between $2\,500$ and $15\,000$ km s$^{-1}$. A systematic decline
of the $T_C$ statistics can be observed beyond $m_{\rm lim}=15.35$. 
Top panel shows in logarithmic scale the galaxies number count in
function of the limiting apparent magnitude $m_*$. The $0.6$ slope
(gray line) is the slope expected if the sources were uniformly
distributed in space.}
\label{figure3}
\end{figure}

The test is applied to subsamples truncated to increasing values
of the limiting apparent magnitude $m_*$. For each galaxies with
coordinates $(M_i,Z_i)$, the quantity $M_{\rm lim}(Z_i)=m_* - Z_i$
is formed, the numbers $r_i$ and $n_i$ are computed, as well as the random
variable ${\hat \zeta}_i$ and its variance $V_i$. The statistic
$T_c(m_*)$ is obtained following Eq. (\ref{T_C}) by summing over
all the galaxies of the subsample. The results are shown figure 3,
bottom panel. Considering the value of $T_C \approx -5$ at $m_*=15.5$,
it is clear that the completeness
in apparent magnitude of the SSRS2 sample is not satisfied up to the 
magnitude $15.5$. Or in other words, the assumption that the
sample is complete up to $15.5$ is rejected at a confidence level
greater than $5\sigma$. The rule for deciding which value of 
the completeness limit 
has to be adopted
is on the other hand a matter of choice. Herein a $2\sigma$ criterion
has been chosen to reject the completeness hypothesis, leading
to a value of $m_{\rm lim}=15.35$ for the completeness in apparent
magnitude. 

The decimal logarithm of the number count versus the limiting apparent magnitude
is shown figure 3, top panel.
A slope of $0.6$ is expected  
if galaxies were uniformly distributed in
space (i.e. this is the standard completeness test proposed in Hubble 1926). 
The test does not allow to draw any firm conclusions concerning the
completeness of the sample. The observed slope appears to be
slightly shallower than $0.6$ but nothing prevents the effect
to be due to inhomegeneity in the spatial distribution of the galaxies.
Moreover, no particular trend is visible beyond the $m_*=15.35$ limit where 
the $T_C$ statistic indicates that the SSRS2 sample suffers from
incompleteness.

The application of the $T_C$ statistic to the SSRS2 sample allows to conclude,
with a high confidence level, that the sample is not complete
in apparent magnitude up to  $m_{\rm SSRS2}=15.5$ mag. However, the 
significant deviation of the $T_C$ statistic from zero may  be
due to hidden systematic effects. In particular, it has been assumed that the
luminosity function does not depend on the spatial position of the galaxies.
It is well known that the E/SO galaxies, in contrast to spirals,
populate preferentially galaxy clusters (see for example Loveday et al. 1995). 
If the luminosity functions of E/SO and spiral galaxies
are indeed different (say for example that E/SO galaxies are brighter 
in average), the luminosity function of the whole population (E/SO$+$spirals)
is expected to depend on the spatial position (in average, galaxies will be 
brighter in clusters than in the field). Such an environmental effect
could influence the $T_C$ statistic and therefore affect the conclusions
drawn from the completeness test. 
The influence of E/S0 and spiral galaxies segregation is investigated 
figure 4. The completeness test has been applied separately to the
two populations. The systematic decline of the $T_C$ statistic can be
observed for both types and the completeness limit of $m_{\rm SSRS2}=15.5$ mag
is ruled out with a high confidence level of rejection ($\approx 4\sigma$).

\begin{figure}
\vbox
{\epsfig{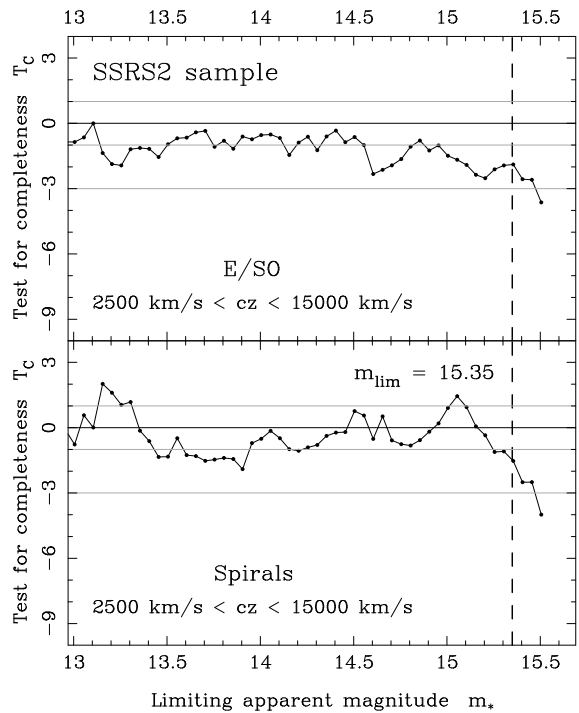}}
\caption{The test for completeness $T_C$ applied to the
$1373$ E/SO galaxies (Top) and to the $2780$ spirals galaxies 
(Bottom) of
the SSRS2 sample with redshifts
between $2\,500$ and $15\,000$ km s$^{-1}$. The systematic decline
at $m_{\rm lim}=15.35$ is present for both E/SO and spirals galaxies.
}
\label{figure4}
\end{figure}

As pointed out by Jon Loveday, the referee of this paper, other
potential problems may affect the present analysis. It has been assumed
for example that the k-correction and the Galactic extinction map used herein
are correct. An improper correction of these quantities could lead
to systematic biases for the $T_C$ statistic. The influence of such
effects is investigated in figure 5. The top panel shows the completeness
test applied to a subsample of all types nearby galaxies
of the SSRS2 survey (  
$2\,500$  km s$^{-1}$ $< cz <$
$7\,500$ km s$^{-1}$).
 The k-correction for this
subsample is therefore small, $0.048$ mag in average with a maximum
of $0.11$ mag. The decline of the $T_C$ statistic 
at $m_{\rm lim}=15.35$ 
is still present, suggesting that an improper k-correction term cannot
explain the imcompleteness observed in figure 3. The bottom panel
of figure 5 shows the completeness test applied to the all types SSRS2
galaxies with a Galactic extinction $A_g(l,b)$ lesser than 0.1 mag.
Note that such a subsampling is not supposed to bias the $T_C$ statistic
since selection effect in angular direction are allowed (see section 2.2.1).
Again, the $T_C$ statistic falls beyond 
$m_{\rm lim}=15.35$. It thus appears that the systematic deviation of
the completeness test is not due to an eventual improper correction 
for the k-correction or for the Galactic extinction. 

Another potential problem is that the k-correction
is herein type dependent, which is not described by the statistical formalism
presented section 2.1, at least if the luminosity function
is also type dependent. To be strictly valid, the completeness test has
to be applied type by type. Six subsamples of spirals have been selected
(from type $T=1$ to $T=6$). The result of the completeness test at
$m_*=15.5$ mag on these subsamples is respectively $T_C=-1.51$, 
$-1.09$, $-2.73$,
$-2.33$, $-0.94$ and $T_C=-1.6$, which indicates 
that in average the completeness
of the sample is not satisfied up to
$m_{\rm lim}=15.5$ mag. Note however that the incompleteness is less flagrant
in this type-by-type analysis
than  for the all spirals subsample of figure 4 ($T_C=-4$ at
$m_*=15.5$ mag). 
It is not surprising since, as any rejection test,
the efficiency of the $T_C$ statistic is improved as the number of 
the datapoints increases.

\begin{figure}
\vbox
{\epsfig{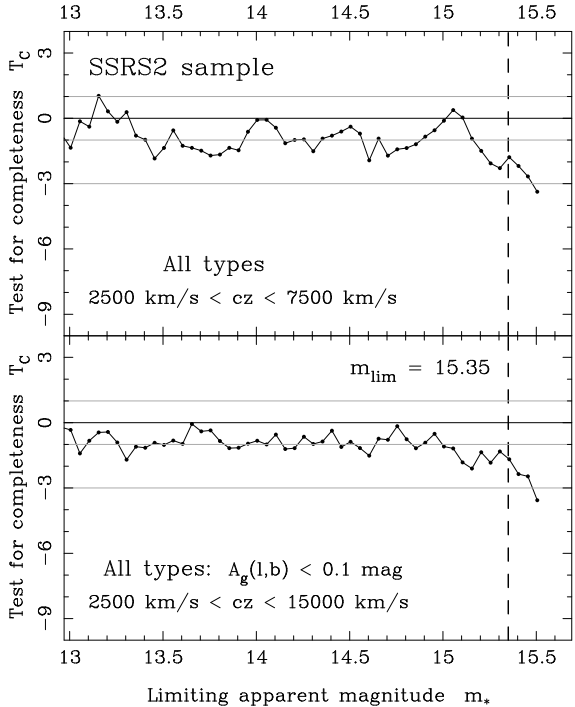}}
\caption{The test for completeness $T_C$ applied to the SSRS2 all types
galaxies
(top) with redshifts
between $2\,500$ and $7\,500$ km s$^{-1}$ ($2416$ galaxies with
a mean k-correction of $0.048$ mag) 
(bottom) with redshifts
between $2\,500$ and $15\,000$ km s$^{-1}$ and a galactic extinction
$A_g(l,b) < 0.1$ mag ($1260$ galaxies). 
The systematic decline
at $m_{\rm lim}=15.35$ is present in both cases, suggesting that the
imcompleteness cannot be explained by an improper
k-correction term or Galactic extinction correction.
}
\label{figure5}
\end{figure}

The results of the test for completeness on the
SSRS2 sample can be summarised as follows. 
The test for completeness indicates that 
the primary list of galaxies
used to build on the SSRS2 redshifts sample is not complete in appparent
magnitude to the $m_{\rm SSRS2}=15.5$ mag limit. Rather, it suggests 
to adopt 
$m_{\rm SSRS2}=15.35$ mag (or less)
as a reasonable completeness limit.
The point is of importance since incompleteness in apparent magnitude
is source of
biases when evaluating
the intrinsic characteristics of the galaxies
luminosity function for example (Marzke et al. 1998).
A special attention has been paid to investigate the systematic effects which
could bias the $T_C$ statistic (influence of peculiar velocities,
evolution of the luminosity function with time, environmental effect,
influences of the k-correction and galactic extinction).  
 
\section{Summary}

A new tool has been  proposed for assessing the completeness
in apparent magnitude
of a magnitude-redshift sample. The technique 
presents a real improvement
compared to standard
completeness tests. Namely, no a priori assumptions are required
concerning the
redshift space distribution of the sources. It means in particular
that neither the clustering nor the evolution of the mean number density
of the galaxies
do affect the result of the search.

The test for completeness presented herein is however
reliable
if and only if the magnitude-redshift sample verifies the following
criteria:
\\{\i})
 The distances of the galaxies are required, which implies
that a cosmological world model has to be specified and that the
contribution of peculiar velocities to observed redshifts can be
safely neglected. Furthermore, the galactic extinction and the k-correction 
involved in the definition of the absolute magnitudes are required. 
\\{\i}{\i})
The shape of the luminosity function of the galaxies is
not allowed to change with time (in practice, this criterion is
achieved if the sample is divided  into thin intervals of redshift, i.e. time).
\\{\i}{\i}{\i})
The luminosity function of the population is assumed to be independent
on the spatial position of the galaxies. In particular, environmental
effect may affect the results of the completeness test.   
 
\section*{Acknowledgements}
I would like to thank Richard Barrett, Martin Hendry, Gilles Theureau 
and David
Valls-Gabaud for fruitful discussions. I acknowledges
the support of the PPARC and the use of the
STARLINK computer node at Glasgow University.

\end{document}